\documentclass[12pt,preprint]{aastex}
\usepackage{psfig,epsfig,amsfonts,amsmath,amssymb,graphicx,morefloats,appendix}
\usepackage{color}
\usepackage{natbib}
\begin{document}

%\shorttitle{Millimeter Emission in HD 15115}
%\shortauthors{MacGregor et al.}

\slugcomment{Accepted by ApJ: January 22, 2015}

\title{
Resolved Millimeter Emission from the HD~15115 Debris Disk
}

\author{
Meredith A. MacGregor\altaffilmark{1},
David J. Wilner\altaffilmark{1},
Sean M. Andrews\altaffilmark{1},
A. Meredith Hughes\altaffilmark{2}
}
\altaffiltext{1}{Harvard-Smithsonian Center for Astrophysics,
  60 Garden Street, Cambridge, MA 02138, USA}
\altaffiltext{2}{Department of Astronomy, Van Vleck Observatory, 
  Wesleyan University, 96 Foss Hill Drive, Middletown, CT 06459, USA)}

\begin{abstract}

We have used the Submillimeter Array (SMA) to make 1.3 millimeter 
observations of the debris disk surrounding HD 15115, an F-type star 
with a putative membership in the $\beta$ Pictoris moving group.  
This nearly edge-on debris disk shows an extreme asymmetry in optical 
scattered light, with an extent almost two times larger to the west of the 
star than to the east (originally dubbed the ``Blue Needle").  The 
SMA observations reveal resolved emission that we model as a 
circumstellar belt of thermal dust emission. This belt extends to a 
radius of $\sim 110$ AU, coincident with the break in the scattered light 
profile convincingly seen on the western side of the disk. This outer edge 
location is consistent with the presence of an underlying population of 
dust-producing planetesimals undergoing a collisional cascade, as 
hypothesized in ``birth ring'' theory.  In addition, the millimeter emission 
shows a $\sim3\sigma$ feature aligned with the asymmetric western 
extension of the scattered light disk. If this millimeter extension is real,
then mechanisms for asymmetry that affect only small grains, such as 
interactions with interstellar gas, are disfavored.  This tentative feature 
might be explained by secular perturbations to grain orbits introduced by 
neutral gas drag, as previously invoked to explain asymmetric morphologies 
of other, similar debris disks.
\end{abstract}


\section{Introduction}

Nearly a hundred dusty debris disks around nearby stars have been spatially 
resolved at one or more wavelengths \citep{mat14}. The bulk radial structure 
of these disks is generally well explained by the presence of a localized belt 
of planetesimals, or ``birth ring,'' where smaller and smaller dust grains 
are produced in a catastrophic collisional cascade and dispersed 
\citep{str06,aug06}.  However, many disks exhibit additional substructure 
such as brightness asymmetries, offsets, warps, and clumps that cannot be 
explained by the steady-state collisional models assumed in this framework.
Numerous mechanisms have been proposed to explain morphologies that depart 
from axial symmetry, including interactions with interstellar gas, planetary 
resonances, and 
stellar flybys \citep[][and references therein]{deb09,man09,man08,kal07}.  Still, 
most of these structures have been detected only at optical and infrared 
wavelengths, probing a population of small grains that react strongly to 
stellar radiation and winds.  Additional observational constraints, 
particularly at millimeter wavelengths, are needed to access larger grains 
with dynamics more similar to the parent colliding bodies, to solidify our 
understanding of the mechanisms that shape debris disk structure.

HD 15115 is an F2V star at $45 \pm 1$ pc \citep{vanL07} whose space motions 
suggest membership in the young $\beta$ Pictoris moving group \citep{moor11}, 
which includes the well-studied $\beta$ Pic and AU Mic debris disks. 
Although the case for membership in this group is not ironclad 
\citep[see][]{deb08}, recent estimates suggest an age of $21\pm4$~Myr for 
these stars \citep{bin14}.
An infrared excess from HD~15115 indicating orbiting dust was first noted 
from \emph{IRAS} observations \citep{sil00}. Subsequent scattered light 
imaging from \emph{HST}, Keck, LBT, and Gemini have resolved a remarkable 
edge-on circumstellar disk \citep{kal07,deb08,rod12,sch14,maz14}. 
This disk shows an extreme asymmetry in optical scattered light \citep{kal07}: 
the east side of the disk extends to $\sim7\arcsec$ (315 AU), while the west 
side reaches 
$>12\arcsec$ ($> 550$ AU). Moreover, there are indications that this asymmetry 
has roots at smaller scales. As described by \citet{deb08}, 
the optical surface brightness 
drops steadily on the east side from $\sim 1\arcsec$ ($< 45$ AU),
while the western side appears to flatten interior to $\sim 2\arcsec$ 
($\sim 90$ AU). In ``birth ring'' theory, this flattening corresponds to the 
outer edge of the planetesimal belt where smaller grains are created in 
collisions and are subsequently launched by stellar forces into wider 
bound and unbound orbits. The presence of a ring structure is supported 
by a subtle concavity visible in scattered light to the north of the star 
\citep{sch14}, and this brighter side of a ring becomes clear in recent 
high contrast images \citep{maz14}.

A $3\sigma$ detection of 850 $\mu$m continuum emission from the HD~15115 
system using the James Clerk Maxwell Telescope/SCUBA suggested the presence 
of a reservoir of large dust grains in the disk, with low temperature 
($T_d = 62$~K) 
and low mass ($\sim 0.047$ M$_\oplus$) indicative of debris \citep{wil06}. 
Subsequent mapping observations with SCUBA-2 obtained a consistent but 
higher flux density ($8.5 \pm 1.2$ mJy vs. $4.9 \pm 1.6$ mJy) and did not 
resolve any structure with a $14''$ (FWHM) beam \citep{pan13}. In light of 
the asymmetric morphology of the HD~15115 disk seen in scattered light, 
we have used the Submillimeter Array (SMA)\footnote{The Submillimeter Array 
is a joint project between the 
Smithsonian Astrophysical Observatory and the Academica Sinica Institue of 
Astronomy and Astrophysics and is funded by the Smithsonian Institution and 
the Academica Sinica.} to resolve the millimeter emission from the disk,
aimed at tracing the distribution of underlying planetesimals.  These 
observations reveal a planetesimal belt together with tentative evidence for 
an asymmetric extension of millimeter emission to the west, coincident with 
the extended feature seen in optical scattered light. We quantify the detected 
millimeter structure and discuss the plausibility of proposed mechanisms for 
the asymmetry.

\section{Observations}

We observed HD 15115 in fall 2013 with the SMA \citep{ho04} on Mauna Kea, 
Hawaii at a wavelength of 1.3 mm using both the compact and extended 
configurations of the array. Table~\ref{tab:obs} summarizes the essentials 
of these observations, including the dates, baseline lengths, and atmospheric 
opacity.  With the extended baselines, these observations probe angular scales 
down to $\lesssim1\arcsec$.  While typically only six (or fewer) of the eight 
array antennas were available during the five tracks, the weather conditions 
were generally very good for observations at this wavelength. 
The total bandwidth available was 8 GHz derived from two sidebands spanning 
$\pm4$ to 8 GHz from the local oscillator (LO) frequency. 
The phase center was located at $\alpha = 02^\text{h}26^\text{m}16\fs24$, 
$\delta = +06\degr17\arcmin33\farcs19$ (J2000), corresponding to the position 
of the star uncorrected for its proper motion of (86.31,-49.97) mas~yr$^{-1}$ 
\citep{vanL07}.  The $\sim 54''$ (FWHM) field of view is set by the primary 
beam size of the 6-m diameter array antennas.  

The data from each track were calibrated independently using the IDL-based 
MIR software package.  Time-dependent complex gains were determined 
using observations of two nearby quasars, J0224+069 ($0\fdg8$ away) and 
J0238+166 ($10\fdg8$ away), interleaved with observations of HD~15115 in a 
12 minute cycle for the extended tracks and a 16 minute cycle for the compact 
tracks. The passband shape was calibrated using available bright sources, 
mainly 3C84 or 3C454.3.  Observations of Uranus during each track were used 
to derive the absolute flux scale with an estimated accuracy of $\sim10\%$.  
Imaging and deconvolution were performed with standard routines in the 
MIRIAD software package. A variety of visibility weighting schemes were
used to explore compromises in imaging between higher angular resolution 
and better surface brightness sensitivity. 

\section{Results and Analysis}
\label{sec:analysis}

\subsection{1.3 mm Dust Continuum Emission}

Figure 1 shows a contour image of the 1.3 millimeter emission overlaid on 
a \emph{Hubble Space Telescope}/Advanced Camera for Surveys coronographic 
image of optical scattered light (F606W filter) from \cite{kal07}.  
The synthesized beam size for this 1.3 millimeter image, obtained with
natural weighting, is $3\farcs0 \times 2\farcs1$ ($135 \times 95$ AU),
position angle $73\degr$.  The rms noise is 0.30~mJy~beam$^{-1}$, and 
the peak signal-to-noise ratio is about 5.
The red star symbol indicates the stellar position, corrected for proper 
motion, i.e. offset from the phase center by $(1\farcs3,-0\farcs7)$. 
This position coincides very closely with the emission peak, well within 
the combined uncertainties of the absolute astrometry and modest 
signal-to-noise.
The image reveals a narrow band of 1.3 millimeter continuum emission that
extends from the east-southeast to west-northwest, aligned closely with 
the optical scattered light disk orientation.
The western extent of the millimeter emission appears to be greater than 
the eastern extent, although the significance of this elongation is not high.

\begin{figure}[ht]
\centerline{\psfig{file=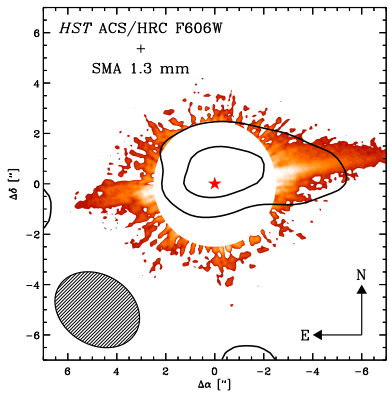,width=6cm,angle=0}}
\caption[]{\small SMA image of the 1.3 millimeter continuum emission from 
HD~15115, overlaid on an image of optical scattered light from the 
\emph{Hubble Space Telescope} \citep{kal07}.  The contour levels are in 
steps of $2 \times 0.3$ mJy, the rms noise level.  The ellipse in the lower 
left corner indicates the $3\farcs0 \times 2\farcs1$ (FWHM) synthesized beam 
size. The star symbol marks the position of the stellar photosphere, corrected
for proper motion. 
}
\label{fig:fig1}
\end{figure}

\subsection{Disk Modeling}
In order to characterize the millimeter emission from HD 15115, we used the 
procedure described by \cite{mac13} that employs Markov Chain Monte Carlo 
(MCMC) methods to fit a simple parametric model to the observed visibilities. 
We assume that the emission arises from a geometrically thin, 
axisymmetric belt characterized by a surface brightness profile of the 
form $I_\nu(r) \propto r^{x}$ for $R_\text{in} < r < R_\text{out}$. 
The belt emission normalization is defined by a total flux density
$F_\text{belt} = \int I_\nu d\Omega$, 
and the belt center is given by offsets relative to the pointing center 
$\{\Delta\alpha,\Delta\delta\}$.
The various scattered light observations show that the disk is nearly 
edge-on to the line of sight.  For our models, we adopt an inclination angle 
of $87\degr$ and an orientation on the sky described by a position angle of 
$278\fdg5 \pm 0.5$ \citep{kal07}. Small variations in these values have
no material impact on the results. This initial model does not attempt to
account for any asymmetric structure in the emission. To address 
the western extension, we also made models that included an unresolved source 
with flux density $F_\text{res}$ and position $\Delta r$, defined as an offset 
from the belt center along the position angle of the disk major axis.

For a given parameter set, we compute synthetic visibilities corresponding
to the SMA observations and compare directly to the data with a $\chi^2$ 
value (the sum of real and imaginary components over all spatial frequencies).
By fitting the visibility data directly, we are not sensitive to the 
non-linear effects of deconvolution, and take full advantage of the 
complete range of spatial frequencies sampled by the observations. 
We assumed uniform priors for all parameters, with reasonable bounds imposed 
to ensure that the model was well-defined: $F_\text{belt} \geq 0$ and 
$0\leq R_\text{in} < R_\text{out}$.  
In addition, we constrained the two parameters $\{\Delta\alpha,\Delta\delta\}$ 
that describe the belt center to be within $0\farcs5$ of the offsets predicted 
from the stellar proper motion; this generously accomodates the uncertainties 
in the proper motion and the absolute astrometry of the observations.
The fit quality is characterized by a likelihood metric, $\mathcal{L}$, 
determined from the $\chi^2$ values (ln$\mathcal{L} = -\chi^2/2$).  
We make use of the affine-invariant ensemble MCMC sampler proposed by 
\cite{goo10} and implemented effectively in \texttt{Python} by \cite{for13}.  
With this MCMC approach, we can characterize efficiently the multidimensional 
parameter space of this simple model and determine posterior probability 
distribution functions for each parameter by marginalizing over all 
other parameters in turn.

\begin{figure}[ht]
\centerline{\psfig{file=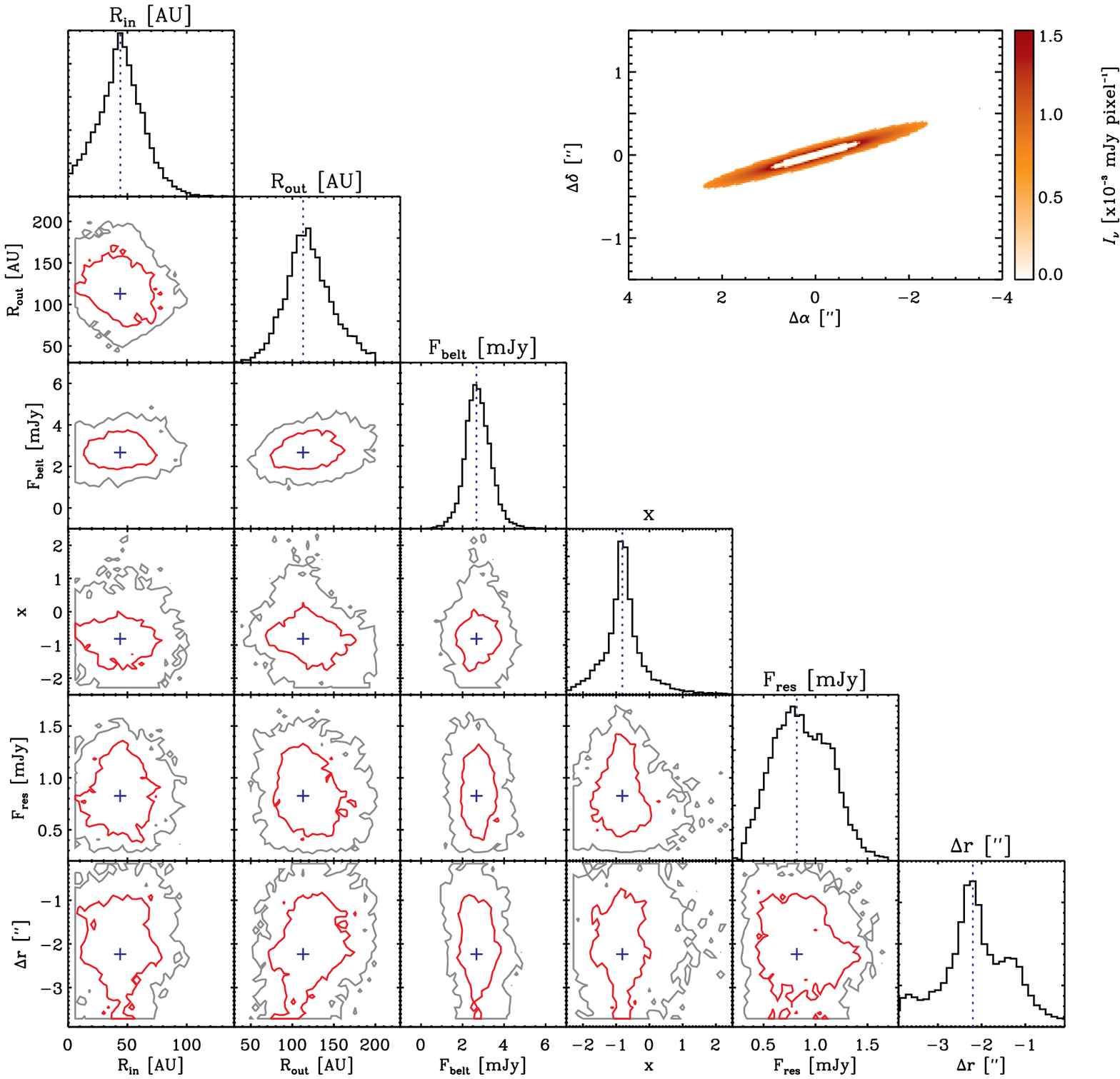,width=14cm,angle=0}}
\caption[]{\small 
A sample of the output from a run of $\sim10^4$ MCMC trials 
for the 6 best-fit model belt parameters ($R_\text{in}$, $R_\text{out}$, 
$F_\text{belt}$, $x$, $F_\text{res}$, and $\Delta r$). The diagonal panels show the 1D histogram for 
each parameter marginalized over all other parameters considered in the model.  For each parameter, 
the peak of each histogram is taken to be the best-fit value.
The remaining panels show contour plots of the $1\sigma$ (red) and $2\sigma$ 
(gray) regions for each pair of parameters, with the blue crosses marking the 
best-fit values.  The inset panel in the upper right shows the resulting 
best-fit model at full resolution, pixel scale $\sim 0\farcs02$ (0.9 AU).  The $1\sigma$ and $2\sigma$ 
regions are determined by assuming normally distributed errors, where the probability that a 
measurement has a distance less than $a$ from the mean value is given by $\text{erf}\left(\frac{a}{\sigma\sqrt{2}}\right)$.
}
\label{fig:fig2}
\end{figure}   

\subsection{Results of Model Fits}

Figure~\ref{fig:fig2} shows the output of $\sim10^4$ MCMC trials and
Table~\ref{tab:models} lists the best-fit model parameter values and their 
$68\%$ uncertainties determined from the marginalized posterior probability 
distributions for the full model (axisymmetric belt plus western extension).  
Figure 3 shows comparisons between the data and the best-fit models in the 
image plane, including the imaged residuals in the rightmost panels. The simple 
symmetric disk model in the upper panels of Figure 3 reproduces the bulk 
of the millimeter emission well, but the residual image shows a $\sim 3\sigma$ 
feature coincident with the western extension of the scattered light along 
the disk axis, and this model yielded a reduced $\chi^2 = 3.67$ 
(153,032 independent data points, 6 free parameters).
The departure from unity reflects the poor fit of the model, in part due 
to the presence of the residual western emission.  By including additional 
parameters in the model that account for this residual feature 
($F_\text{res}$ and $\Delta r$), the model fit was more satisfactory, as 
can be seen in the lower panels of Figure 3, with reduced $\chi^2 = 1.88$.

\begin{figure}[htbp]
\centerline{\psfig{file=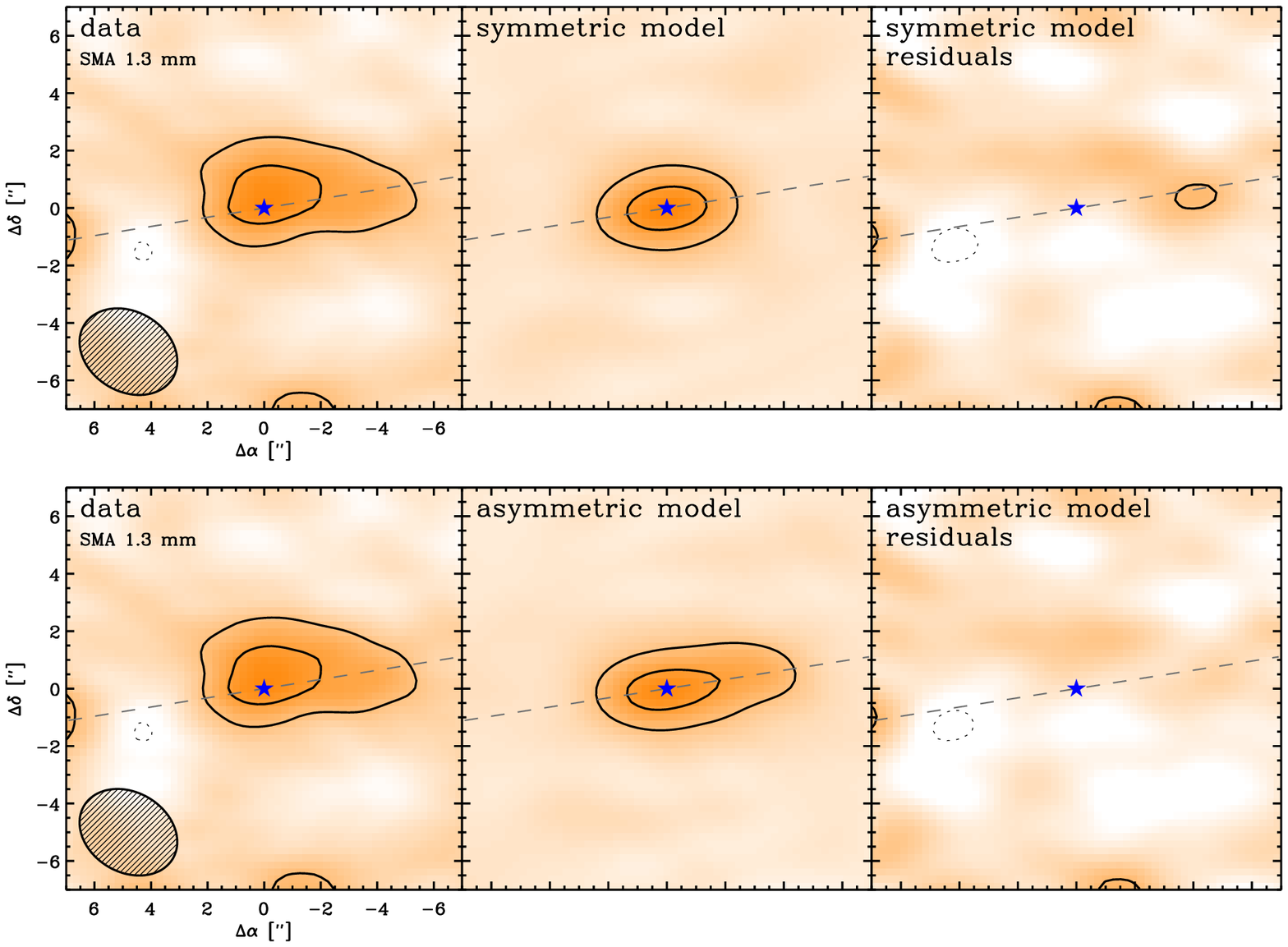,width=15cm,angle=0}}
\caption[]{\small 
\emph{(upper): (left)} The 1.3 millimeter continuum emission 
from HD 15115 observed with the SMA, as in Figure~\ref{fig:fig1},
\emph{(center)} image of the best-fit azimuthally 
symmetric disk model (see Section~\ref{sec:analysis} for a description of the modeling 
formalism and results), and \emph{(right)} the imaged residuals from the symmetric 
model, showing a $3\sigma$ feature on the western side of the disk. 
\emph{(lower): (left)} the SMA image, \emph{(center)} image of the best-fit asymmetric
model, and \emph{(right)} the imaged residuals from the asymmetric model, which
do not show any remaining significant features. The contour levels are at
$2\sigma$ (0.6 mJy~beam$^{-1}$) intervals in all panels.
}
\label{fig:fig3}
\end{figure}

The total flux density of the best fit model is constrained to be
$F_\text{belt} = 2.6^{+0.5}_{-0.8}$ mJy. If we extrapolate the 
available submillimeter measurements in the literature using the typical spectral index of $\sim 2.65$ for debris disks at these wavelengths \citep{gas12}, we find close agreement with the SMA value.  In particular, extrapolating the SCUBA-2 observation of \citet{pan13} 
to 1.3 mm gives $2.8 \pm 0.4$ mJy, while the older SCUBA observation of
\citet{wil06} gives $1.6 \pm 0.5$ mJy.  Given this consistency, it seems 
likely that the SMA observations detect the full disk emission and do not
miss any significant, more extended, millimeter flux.

The location of the outer edge of the modelled millimeter emission belt, 
$R_\text{out} = 110^{+31}_{-22}$~AU, can be compared to expectations 
based on the scattered light radial surface brightness profile and
``birth ring'' theory.
The \emph{HST} imaging of HD~15115 by \citet{deb08} shows that the 
scattered light profile steepens on the western side of the disk,
falling as $r^{-3.56 \pm 0.06}$ beyond $1\farcs8$ (81 AU).  
The ground based scattered light observations of \citet{rod12} show 
a similar break in the western surface brightness profile near 
$\sim1.8- 2\arcsec$ ($\sim 80 - 90$ AU). The model fit to the outer edge of 
the millimeter belt is consistent with both of these determinations, 
within the substantial uncertainties. However, in the context of theory, 
it remains unclear why the eastern side of the scattered light disk shows 
no clear break in surface brightness slope.

The observations of \citet{rod12} additionally show a drop in the 
scattered light surface brightness of the disk's western side \emph{interior} 
to $1\arcsec$ (45~AU), presumably marking the inner edge of the belt. 
The higher contrast images of \citet{sch14} and \citet{maz14} confirm this 
feature and suggest at least a partial clearing within this radius. Analysis 
of the spectral 
energy distribution of the system by \citet{moor11} yields a best-fit model 
with an inner warm component and an outer cold ring at a minimum radius of 
$42 \pm 2$ AU.  Both of these determinations of the inner disk edge match 
the (poorly constrained) best-fit inner radius of the millimeter belt, 
$R_\text{in} = 43 \pm 28$ AU.  

Despite the modest signal-to-noise ratio of the millimeter observations, 
this model fitting exercise yields constraints on the inner and outer radii 
of the millimeter emission belt that are consistent with previous 
observations and compatible with the ``birth ring'' model of debris disks.

Not surprisingly, given the limits of the resolution and sensitivity of
the SMA data, model fitting does not provide a strong constraint on the 
power-law gradient
of the millimeter radial emission profile, $x=-0.75^{+1.37}_{-0.87}$.  
The well known degeneracy between the emission gradient, $x$, and disk outer 
radius, $R_\text{out}$, \citep[e.g.][]{mun96} is apparent in 
Figure~\ref{fig:fig2} as a slope introduced in the lower contours resulting 
from that pair of model parameters. 
For very negative values of $x$, the contours spread to span a wide range of 
potential outer radii, as the width and extent of the emission belt become 
difficult to constrain due to the low brightness in the outer regions. 
Similarly, the inner radius of the disk is poorly constrained for positive 
gradients.  If we assume that the emitting dust is in radiative equilibrium 
with stellar heating, leading to a temperature gradient described by 
$T \propto r^{-0.5}$, then taking the best fit exponent at face value implies 
a relatively shallow surface density profile, $\Sigma \propto r^{-0.25}$ 
(albeit with large uncertainty).

\section{Discussion}

Millimeter imaging of HD~15115, like its suspected sister stars in the 
$\beta$ Pictoris moving group, $\beta$ Pic \citep{wil11} and AU Mic 
\citep{wil12,mac13}, shows a resolved belt of emission with outer edge 
coincident with a previously observed break in scattered light, consistent 
with an underlying ``birth ring'' of colliding planetesimals.  
For all of these stars, resolved multi-wavelength datasets suggest
planetesimal collisions within a belt produce grains with a spectrum 
of sizes, and the effects of size-dependent dust dynamics produce the 
compact millimeter emission and an extended scattered light halo.  
HD 181327, a fourth member of the $\beta$ Pictoris moving group, also 
shows a belt of cold dust seen in \emph{Herschel} PACS 70 $\mu$m and 
ATCA 3.2 mm images \citep{leb12}, though higher resolution millimeter images 
are still needed to determine its underlying structure.

The detection of a $\sim 3\sigma$ residual along the western extent 
of the HD~15115 disk, aligned with the asymmetry seen in scattered light 
from small grains, provides tantalizing evidence that the distribution of 
larger grains in this system may be asymmetric as well. Similar asymmetric 
debris disk morphologies in scattered light have been observed in other highly 
inclined systems where the geometry favors detection, notably HD 32297 
\citep{cur12,deb09,boc12}
and HD 61005 \citep{hin07,man09}, as well as $\beta$~Pic \citep{kal95}.  
While all of these disks show unique morphological details, similar physical 
mechanisms may be responsible for the overall shaping of the asymmetric 
appearance of the debris. However, it is unclear if the asymmetries can be 
clearly traced back to the dust-producing parent bodies. The constraints from 
millimeter observations so far fail to provide a consistent picture. HD 32297 
was observed at 1.3 millimeters in the early days of CARMA \citep{man08}, and 
the millimeter image shows a hint of asymmetry, albeit at low signal-to-noise.  
On the other hand, SMA 1.3 millimeter observations of HD 61005 \citep{ric13} 
provide no evidence that the millimeter grains are present in the spectacular 
swept-back wings seen in scattered light.  

The most common mechanisms invoked to explain the scattered light asymmetries 
in these highly inclined debris disks involve interactions with the local 
interstellar medium (ISM).  Ram pressure from interstellar gas can remove 
bound and unbound grains from a disk directly, or by inducing outflows of disk 
gas that entrain the grains \citep{man09,deb09}.  Additionally, neutral gas 
can introduce secular perturbations to the orbits of bound grains, producing 
significant effects on disk morphology on orbital timescales \citep{man09}.  
Alternatively planets within the system can have dust-trapping resonances 
that create clumpy, wavelength-dependent grain distributions \citep{wya06}.
Looking outside of the system entirely, external perturbation by a close 
stellar flyby can also shape the distribution of disk grains \citep{kal07}.  
An increase in the dust grain scattering cross section or a local 
density enhancement could produce a local increase of collisions \citep{maz14}.
Finally, azimuthal asymmetries could be generated through the ``photoelectric
instability'' effect, if the disk has a high enough total dust to gas mass ratio \citep{lyr13}.
Each of these mechanisms affects different sized grains preferentially and 
works under specific physical conditions.  In the following, we examine
the plausibility of each of these mechanisms for creating asymmetric 
structures in the HD 15115 disk and similar systems.

\subsection{Interactions with the ISM}

Because the proper motion of HD 15115 is nearly parallel to the disk major 
axis, \citet{deb09} suggest that the dominant disk asymmetry in scattered 
light arises from an ongoing interaction with the local ISM.  In this picture, 
the eastern side of the disk plows head first into the ISM, causing that side 
to become truncated, while small grains are blown out to the west.  
Motivated by the swept back morphologies of HD 32297, HD 15115, and HD 61005, 
\citet{deb09} construct a model for a disk interacting with the gaseous ISM,
based on the gas drag felt by meteorites entering the Earth's atmosphere.
This results in a scaling law for the radial distance ($R_\text{deflect}$) 
at which unbound grains are significantly perturbed from their original 
circumstellar orbits:

\begin{equation}
R_\text{deflect} = 286\left(\frac{1.67\times10^{-22}\text{ g cm}^{-3}}{\rho_\text{gas}}\right)^\frac{1}{2}\left(\frac{20\text{ km s}^{-1}}{v_\text{rel}}\right)\left(\frac{L_\star}{L_\odot}\right)^\frac{1}{2}\text{ AU},
\end{equation}

\noindent where $\rho_\text{gas}$ is the ISM gas density and $v_\text{rel}$ 
is the relative cloud-disk velocity.  By applying the definition of 
$\beta$, the ratio of the force of radiation pressure to the gravitational 
force exerted by the star, Equation (1) can be recast into a scaling relation 
for the ISM gas density ($n_\text{gas}$) and disk relative velocity ($v_\text{rel}$) needed to perturb unbound grains from their orbits.
Assuming perfectly absorbing spherical grains \citep{bur79}

\begin{equation}
\beta = 0.574\left(\frac{L_\star}{L_\odot}\right)\left(\frac{M_\star}{M_\odot}\right)^{-1}\left(\frac{1 \text{ g cm}^{-3}}{\rho_\text{dust}}\right)\left(\frac{1\text{ }\mu\text{m}}{r_\text{dust}}\right),
\end{equation}

\noindent which yields the relation 

\begin{eqnarray}
\left(\frac{n_\text{gas}}{200\text{ cm}^{-3}}\right)\left(\frac{v_\text{rel}}{30\text{ km s}^{-1}}\right)^2 & \ga & 0.6262\left(\frac{M_\star}{0.95\text{ }M_\odot}\right)\left(\frac{a_\text{dust}}{0.1 \text{ }\mu\text{m}}\right) \nonumber \\
& &\times\left(\frac{\rho_\text{dust}}{2 \text{ g cm}^{-3}}\right)\left(\frac{70 \text{ AU}}{r}\right)^2.
\label{eqn:debes}
\end{eqnarray}

Interpreting \emph{HST} images of HD 61005, \cite{man09} discuss a second model
for ISM interaction in which an interstellar gas cloud removes both bound and 
unbound grains from a disk due to ram pressure stripping.  In this scenario, 
the drag force on a grain must be comparable to or exceed the gravitational 
force binding the grain to the star.  For the bound case, \citet{man09} show
that this leads to the following scaling relation for the cloud density 
($n_\text{gas}$) and the relative cloud-disk velocity (again $v_\text{rel}$)

\begin{eqnarray}
\left(\frac{n_\text{gas}}{200\text{ cm}^{-3}}\right)\left(\frac{v_\text{rel}}{30\text{ km s}^{-1}}\right)^2 &\ga& \left(\frac{M_\star}{0.95 \text{ }M_\odot}\right)\left(\frac{a_\text{dust}}{0.1 \text{ }\mu\text{m}}\right) \nonumber \\
& &\times\left(\frac{\rho_\text{dust}}{2 \text{ g cm}^{-3}}\right)\left(\frac{70 \text{ AU}}{r}\right)^2.
\label{eqn:maness}
\end{eqnarray}

\noindent A similar relation is obtained for the unbound case with 
an additional constant factor accounting for the relative distance traveled 
by a grain parallel and perpendicular to the disk midplane.  Although 
formulated somewhat differently from \citet{deb09}, the resulting expression
is functionally identical with the exception of a constant factor
(compare equations \ref{eqn:debes} and \ref{eqn:maness}).

While ram pressure stripping of disk grains by the ISM seems like an 
attractive explanation for the asymmetry seen in systems like HD 15115,
\cite{mar11} use numerical modeling to show that even grains just above the 
disk blow-out size ($1 - 10 \text{ }\mu$m) are minimally affected by ISM 
interactions and are lost from the disk before they acquire any significant 
inclination.  The millimeter grains, with much smaller values of $\beta$, 
should feel little to no effect from ISM gas. Thus, this argument has 
difficulty explaining an asymmetry in the millimeter disk emission.  
Moreover, \cite{man09} point out that the ISM densities required are 
characteristic of cold clouds ($n\sim50$ cm$^{-3}$, $T\sim20$ K) that 
occupy only a very small volumetric filling factor within the Local Bubble.

\cite{man09} also propose that disk \emph{gas} could undergo ram pressure 
stripping by the ISM, and disk grains are consequently swept away when they 
become entrained in this outflowing gas.  
In this scenario, sufficient disk gas is required and the grains must be 
entrained effectively, neither of which is assured. 
However, measurements of the NA I doublet towards HD 15115 by \citet{red07} 
indicate that the column density for this disk is $\sim5$ times 
greater than towards $\beta$ Pic; he estimates that the upper limit on the 
total gas mass in the circumstellar disk is $\sim 0.3 M_\oplus$.

In order for direct ram pressure stripping of disk grains to occur, the force of the ISM 
on the grains must be comparable to the stellar gravitational force.  
Yet, \cite{man09} propose that in the case where this condition is not met, 
neutral gas can still affect disk morphology over timescales of 
$\sim 10^3 - 10^4$ years by introducing secular perturbations to the orbits 
of bound grains.  A similar mechanism has been proposed as the removal 
mechanism for dust from our Solar System at $20 - 100$ AU \citep{sch00}.  
Like drag from the Solar wind, neutral gas drag involves the transfer of 
momentum from incident gas particles to grain surfaces.  However, unlike the 
Solar wind, interstellar gas drag tends to increase grain eccentricities 
and semi-major axes, and can eventually unbind grains from the system.  
The normalcy of the interstellar densities, 
velocities, and cloud sizes required by this neutral gas drag model make 
it a promising mechanism for producing the structures in this family of
asymmetrical debris disks.  Furthermore, the presence of detectable gas 
in the disk lends additional credibility to the argument that neutral gas drag 
may play a role in shaping the disk morphology.

\subsection{Interactions with Other Perturbers}

An alternative mechanism to ISM interactions that can produce disk asymmetries 
involves planet induced resonances.  Parent planetesimals locked in resonance 
with orbiting planets can produce large grains that stay in resonance, or, 
if collisions are unimportant, grains can drift into resonances due to 
Poynting-Robertson drag \citep{kri07}.  \cite{wya06} predict that in the 
former case, an asymmetry can be present at short and long wavelengths, 
but absent at intermediate wavelengths.  In this model, the large grains 
that dominate millimeter emission trace the parent planetesimals and thus 
exhibit the same clumpy, resonant distribution.  The small grains dominant 
at short wavelengths are preferentially created in the high-density 
resonant clumps and expelled from the system on short timescales. 
The mid-sized grains traced by intermediate wavelengths remain bound, but 
fall out of resonance due to radiation pressure and are scattered into an 
axisymmetric distribution.  \cite{man08} invoke this prediction to explain 
the multi-wavelength observations of HD 32297, which appears asymmetric at 
both optical and millimeter wavelengths, but symmetric in the mid-infrared. 
A similar picture may be hinted at in the case of HD 15115.  The scattered 
light observations are clearly asymmetric, and the SMA observations suggest
a potential millimeter asymmetry.  However, observations at $3-5\text{ }\mu$m 
show a symmetric disk morphology \citep{rod12}. Detailed dynamical models of 
these systems are needed to draw firm conclusions about this speculation.

Global disk asymmetries could also be produced by a single or periodic 
stellar flyby, as proposed to explain the large scale scattered light
asymmetry of the $\beta$~Pic debris disk \citep{kal95,kal01}.
\citet{kal07} point out that the $\beta$ Pic moving group member, HIP 12545, 
nearby in the sky to HD~15115 \citep[$3\fdg9$][]{moo06}, could be involved
in a past interaction, as these two stars show identical proper motion, 
galactic space motion, and heliocentric distance within the measured 
uncertainties. Morever, the present location of HIP 12545 is in the 
direction of the truncated eastern side of the HD 15115 disk, as in the 
simulations of \citep{lar01} flyby interactions.  However, there are no
indications other than this circumstantial evidence for any interaction.  
Furthermore, the membership of HD 15115 in the $\beta$ Pic moving group has 
been called into question. \cite{moo06} used kinematic arguments to propose 
this group membership. The recent Bayesian analysis by \citet{mal13} suggests 
a higher probability of membership in the Columba Association.  A better 
understanding of the group membership and age of HD 15115 certainly would be 
helpful in unraveling the origins of the debris disk morphology. 

\subsection{Interactions within Disk Material}

Gemini Near-Infrared Coronographic Imager (NICI) data in the H and Ks bands 
\citep{maz14} resolved the HD 15115 disk and detected ansae on both sides at a radius of 
$1.99\arcsec$ ($\sim90$ AU).  
These new observations indicate a ring-like shape with an inner cavity that 
appears symmetric, in contrast with the east-west brightness asymmetry seen in 
other images.  If the inner ring is truly symmetric, an alternative or additional mechanism
for the observed brightness asymmetry is possible.  \cite{maz14} suggest
that both a local increase in the dust grain scattering cross section and/or a 
local density enhancement of small grains could lead toward an increase in 
collisions at the location of the western brightness peak.  A similar 
explanation was recently invoked to explain the millimeter emission clump 
seen in ALMA observations of the $\beta$ Pictoris disk \citep{dent14}. 

We can crudely estimate the mass of small grains needed to account 
{\em entirely} for the tentative excess millimeter emission on the 
western side of the disk. Optically thin dust emission at a temperature, 
$T_\text{dust}$, has flux 
$F_\text{dust}~\approx~\kappa_\nu B_\nu(T_\text{dust})M_\text{dust}/d^2$, 
where $\kappa_\nu$ is the dust opacity, $B_\nu$ is the Planck function, 
$M_\text{dust}$ is the dust mass, and $d = 45$ pc.  
We estimate a characteristic temperature for dust in this excess feature 
of $\sim30$ K, using its approximate radial location ($\sim 160$ AU) 
and assuming radiative equilibrium of starlight with blackbody grains
\citep{bac93}.
To estimate an appropriate dust opacity, we assume that the grain 
size distribution is described by a power-law between 
0.1 $\mu$m and some maximum size $a_\text{max}$, $n(a) \propto a^q$.  
For grains with composition volume fractions of 0.07, 0.21, 0.42, and 0.30 
for silicate, carbon, ice, and vacuum, respectively, \cite{ric10} calculate 
dust opacities as a function of $a_\text{max}$. (Note, the results are not very
sensitive to other reasonable assumptions about composition.) 
Adopting a typical power-law exponent $q=-3.5$, and maximum grain size 
$a_\text{max} = 0.1$ millimeter, models yield a dust opacity, 
$\kappa_\nu \sim 3$ cm$^2$ g$^{-1}$.  
Given these assumptions, the mass of small grains required to account 
for the excess millimeter flux is $\sim 1\times10^{26}$ g, or $\sim 1.4$ times 
the lunar mass.  As this large mass indicates, a very significant collision 
would be required for small grains to account entirely for the western 
millimeter emission feature. Furthermore, such a collisional feature is 
transient and can persist for only a small fraction of the system age
\citep{jac14}.

\cite{lyr13} discuss an additional mechanism by which disk gas can generate azimuthal asymmetries through the ``photoelectric instability'' effect.  In this scenario, a dense region of dust heats the gas through photoelectric heating and creates a local pressure maximum, which in turn attracts additional dust.  The resulting instability could lead to the development of rings, spirals, or other structures within the disk.  \cite{lyr13} suggest that this effect will occur in disks with a total disk dust to gas mass ratio, $\epsilon \sim 1$.  In the case of HD 15115, estimates of the total dust and gas mass in the disk suggest that $\epsilon > 0.16$, in the regime where this mechanism could work.  However, without further analysis, it is unclear whether the level of azimuthal asymmetries produced would be large enough to account for these new observations.

In short, there is still no clear consensus as to which of these mechanisms 
is responsible for shaping the debris around HD 15115 and other, similar 
systems (notably HD 61005 and HD 32297).  Each mechanism affects different 
sized grains preferentially and operates under specific physical conditions. 
Improved constraints on the detailed morphologies of these systems are 
required to reach definitive conclusions.

\section{Conclusions}

We present SMA 1.3 millimeter observations that resolve the emission from
the HD 15115 debris disk, the first at such a long wavelength.  The bulk of the 
millimeter emission is described well by a symmetric belt, consistent with 
the presence of a ``birth ring'' of collisional planetesimals indicated 
by previous observations of scattered light.  The SMA observations also show 
a $\sim3\sigma$ residual feature coincident with the western extension of the 
scattered light along the disk major axis. If real, this additional feature 
hints that the distribution of larger grains may be asymmetric as well.  

There is no consensus concerning the physical mechanism or mechanisms 
responsible for the scattered light asymmetries in HD~15115 and other,
similar debris disks. 
Ram pressure from surrounding interstellar gas can strip both bound 
and unbound grains, or induce outflows of disk gas that entrain grains.
However, the tentative millimeter emission asymmetry disfavors any process
that affects only small grains.  An alternative mechanism invokes neutral gas 
in the disk that introduces secular perturbations to grain orbits, producing 
significant effects on disk morphology on orbital timescales \citep{man09}. 
The presence of atomic gas towards HD~15115 suggests that this idea should be 
explored more thoroughly. Interactions with perturbers are another possibility.
Dust-trapping resonances from orbiting planets may be able to explain the 
observed wavelength-dependent structure.  Future observations of 
HD~15115 at higher angular resolution and sensitivity with the 
Atacama Large Millimeter/Submillimeter Array are needed to resolve the 
detailed morphology of the millimeter emission and to address the processes 
responsible for shaping this remarkable debris disk.

\acknowledgements
We thank Paul Kalas for providing the \emph{Hubble Space Telescope} image 
in Figure 1. M.A.M acknowledges support from a National Science Foundation 
Graduate Research Fellowship (DGE1144152).
We thank Katherine Rosenfeld, Til Birnstiel and John Debes for helpful conversations.

\bibliography{References}

\begin{deluxetable}{c c c c c c c}
\tablecolumns{7}
\tabcolsep0.1in\footnotesize
\tabletypesize{\small}
\tablewidth{0pt}
\tablecaption{Submillimeter Array Observations of HD 15115
\label{tab:steps}}
\tablehead{
\colhead{Observation} & 
\colhead{Array} & 
\colhead{Number of} & 
\colhead{Baseline} & 
\colhead{LO Freq.} &
\colhead{HA} &
\colhead{225 GHz atm.} \\
\colhead{Date} & 
\colhead{Config.} & 
\colhead{Antennas} & 
\colhead{Lengths (m)} & 
\colhead{(GHz)} &
\colhead{Range} &
\colhead{Opacity$^\text{a}$}
}
\startdata
2013 Sep 15 & Extended & 6& $25-174$ & 225.5 & $-4.2, 3.3$ &  $0.08$\\
2013 Oct 19 & Extended  & 5 & $20 - 182$ & 235.5 & $-3.8, 4.9$ & $0.28$  \\
2013 Nov 18 & Extended & 6 & $23 - 174$ & 225.5 & $-2.3, 3.1$ & $0.10$  \\
2013 Dec 10 & Compact & 6 & $8 - 53$ & 225.5 & $-1.3, 3.7$ & $0.20$ \\
2013 Dec 12 & Compact & 6 & $6 - 53$ & 224.9 & $-4.5, 4.8$ & $0.11$
\enddata
\tablecomments{$^\text{a}$ characteristic value for the track measured at 
the nearby Caltech Submillimeter Observatory}
\label{tab:obs}
\end{deluxetable}

\begin{deluxetable}{l l c c}
\tablecolumns{4}
\tabcolsep0.2in\footnotesize
\tabletypesize{\small}
\tablewidth{0pt}
\tablecaption{Model Parameters
\label{tab:steps}}
\tablehead{
\colhead{Parameter} & 
\colhead{Description} & 
\colhead{Best-Fit} & 
\colhead{68\% Confidence Interval} 
}
\startdata
$F_\text{belt}$ & Belt flux density (mJy) & $2.56$ & $+0.50, -0.83$ \\
$R_\text{in}$ & Belt inner radius (AU) & $43.4$ & $+28.3, -28.3$ \\
$R_\text{out}$ & Belt outer radius (AU) & $113.$ & $+30.6, -21.8$ \\
$x$ & Belt radial power law index & $-0.75$ & $+1.37, -0.87$\\
\hline
$F_\text{res}$ & Point source flux (mJy) & $0.84$ & $+0.16, -0.12$ \\
$\Delta r$ & Point source offset ($\arcsec$) & $-3.55$ & $+0.45, -0.12$ \\
\hline
$\Delta\alpha$ & R.A. offset of belt center ($\arcsec$) & $1.26$ & $+0.07, -0.05$ \\
$\Delta\delta$ & Decl. offset of belt center ($\arcsec$) & $-0.78$ & $+0.09, -0.05$
\enddata
\label{tab:models}
\end{deluxetable}

\end{document}